\documentclass{amsart}
\usepackage{amsmath}
\usepackage{amssymb}

\theoremstyle{plain}
\newtheorem{theorem}{Theorem}[section]
\newtheorem{lemma}[theorem]{Lemma}

\theoremstyle{definition}
\newtheorem{definition}[theorem]{Definition}

\theoremstyle{remark}

\numberwithin{equation}{section}

\input{tcilatex}

\begin{document}
\title[Delsarte transmutation operators]{Differential-geometric and
topological structure of multidimensional Delsarte transmutation operators}
\thanks{The third author was supported in part by a local AGH grant.}
\author{Yarema Prykarpatsky*)}
\address{*)The AMM University of Science and Technology, Department of
Applied Mathematics, Krakow 30059 Poland, and Brookhaven Nat. Lab., CDIC,
Upton, NY, 11973 USA}
\email{yarchyk@imath.kiev.ua, yarpry@bnl.gov}
\author{Anatoliy Samoilenko**)}
\address{**)The Institute of Mathematics, NAS, Kyiv 01601, Ukraine}
\author{Anatoliy K. Prykarpatsky***)}
\address{***)The AMM University of Science and Technology, Department of
Applied Mathematics, Krakow 30059 Poland, and Dept. of \ Nonlinear
Mathematical Analysis at IAPMM, NAS of Ukraine, Lviv 01601 Ukraina}
\email{pryk.anat@ua.fm, prykanat@cybergal.com}
\subjclass{Primary 34A30, 34B05 Secondary 34B15 }
\keywords{Delsarte transmutation operators, Darboux transformations,
differential geometric and topological structure, cohomology properties, De
Rham -Hodge-Skrypnik complexes}
\date{November 26, 2003}

\begin{abstract}
A differential geometrical and topological structure of Delsarte
transmutation operators in multidimension is studied, the relationships with
De Rham-Hodge-Skrypnik theory of generalized differential complexes is
stated.
\end{abstract}

\maketitle


\section{Introduction}

\setcounter{equation}{0}Consider the Hilbert space $\mathcal{H}=L_{2}(%
\mathbb{R}^{m};\mathbb{C}^{N})),$ $m,N\in \mathbb{Z}_{+},$ with the scalar
semi-linear form on $\mathcal{H}^{\ast }\times \mathcal{H}$
\begin{equation}
(<\varphi ,\psi >):=\int_{\mathbb{R}^{m}}\bar{\varphi}(x)^{\intercal }\psi
(x)dx  \label{0.1}
\end{equation}%
for any pair $(\varphi ,\psi )\in \mathcal{H}^{\ast }\times \mathcal{H},$
where, evidently $\mathcal{H}^{\ast }\simeq \mathcal{H},$ sign $"\intercal "$%
is the usual matrix transposition. Take also $\mathcal{H}_{0}$ and $\mathcal{%
\tilde{H}}_{0}$ being some two closed subspaces of $\mathcal{H}$ and
correspondingly two linear operators $L$ and $\tilde{L}$ acting from $%
\mathcal{H}$ into $\mathcal{H}.$

\begin{definition}
(J. Delsarte and J. Lions \cite{DL}) A linear invertible operator $\mathbf{%
\Omega }$ \ defined on the whole $\mathcal{H}$ \ and acting from $\mathcal{H}%
_{0}$ onto $\mathcal{\tilde{H}}_{0}$ is called a Delsarte transmutation
operator for a pair of linear operators $L$ and $\tilde{L}:$ $\mathcal{H}%
\rightarrow \mathcal{H}$ if the following two conditions hold: \ \ \ \ \ \ \
\ \ \ \ \ \ \ \ \ \ \ \ \ \ \ \ \ \ \ \ \ \ \ \ \ \ \ \ \ \ \ \ \ \ \ \ \ \
\ \ \ \ \ \ \ \ \ \ \ \ \ \ \ \ \ \ \ \ \ \ \ \ \ \ \ \ \ \ \ \ \ \ \ \ \ \
\ \ \ \ \ \ \ \ \

\begin{itemize}
\item the operator $\mathbf{\Omega }$ and its inverse $\mathbf{\Omega }^{-1}$
are continuos in $\mathcal{H};$ \ \

\item the operator identity
\begin{equation}
\tilde{L}\mathbf{\Omega }=\mathbf{\Omega }L  \label{0.2}
\end{equation}%
is satisfied.
\end{itemize}
\end{definition}

Such transmutation operators were for the first time introduced in \cite{De,
DL} for the case of one-dimensional second order differential operators. In
particular, for the Sturm-Liouville and Dirac operators the complete
structure of the corresponding Delsarte transmutation operators was
described in \cite{Ma, LS}, where also the extensive applications to
spectral theory were given.

As there was become clear just recently, some special cases of the Delsarte
transmutation operators were constructed much before by Darboux and Crum
(see \cite{MS}). A special generalization of the Delsarte-operators for the
two-dimensional Dirac operators was done for the first time in \cite{Ni},
where its applications to inverse spectral theory and solving some nonlinear
two-dimensional evolution equations were also presented.

Recently some progress in this direction was made in \cite{SPS,SP} due to
analyzing a special operator structure of Darboux type transformations which
appeared in \cite{Nim}.

In this work we give in some sense a complete description of
multi-dimensional Delsarte transmutation operators based on a natural
generalization of the differential-geometric approach devised in \cite{SP},
and discuss how one can apply these operators to studying spectral
properties of linear multi-dimensional \ differential operators.

\section{The differential-geometric structure of a generalized Lagrangian
identity}

\setcounter{equation}{0}Let a multi-dimensional linear differential operator
$L:\mathcal{H\rightarrow H}$ of order $n(L)\in \mathbb{Z}_{+}$ be of the
form
\begin{equation}
L:=\sum_{|\alpha |=0}^{n(L)}a_{\alpha }(x)\frac{\partial ^{|\alpha |}}{%
\partial x^{\alpha }},  \label{1.1}
\end{equation}%
where, as usually, $\alpha \in \mathbb{Z}_{+}^{m}$ is a multi-index, $x\in
\mathbb{R}^{m},$ and for brevity one assumes that coefficients $a_{\alpha
}\in \mathcal{S(}\mathbb{R}^{m};End\mathbb{C}^{N}).$ Consider the following
easily derivable generalized Lagrangian identity for the differential
expression (\ref{1.1}) :

\begin{equation}
<L^{\ast }\varphi ,\psi >-<\varphi ,L\psi >=\sum_{i=1}^{m}(-1)^{i+1}\frac{%
\partial }{\partial x_{i}}Z_{i}[\varphi ,\psi ],  \label{1.2}
\end{equation}%
where $(\varphi ,\psi )\in \mathcal{H}^{\ast }\times \mathcal{H},$ mappings $%
Z_{i}:\mathcal{H}^{\ast }\times \mathcal{H}\rightarrow \mathbb{C},$ $i=%
\overline{1,m},$ are semilinear due to the construction and $L^{\ast }:%
\mathcal{H}^{\ast }\rightarrow \mathcal{H}^{\ast }$ is the corresponding
formally conjugated to (\ref{1.1}) differential expression, that is
\begin{equation}
L^{\ast }:=\sum_{|\alpha |=0}^{n(L)}(-1)^{|\alpha |}\frac{\partial ^{|\alpha
|}}{\partial x^{\alpha }}\cdot \bar{a}_{\alpha }^{\intercal }(x).
\label{1.3}
\end{equation}%
Having multiplied the identity (\ref{1.2}) by the usual oriented Lebesgue
measure $dx=\wedge _{j=\overrightarrow{1,m}}dx_{j},$ we get that $\ \ $%
\begin{equation*}
<L^{\ast }\varphi ,\psi >dx-<\varphi ,L\psi >dx=dZ^{(m-1)}[\varphi ,\psi ]
\end{equation*}%
\ for all $(\varphi ,\psi )\in \mathcal{H}^{\ast }\times \mathcal{H},$ where
\begin{equation}
Z^{(m-1)}[\varphi ,\psi ]:=\sum_{i=1}^{m}dx_{1}\wedge dx_{2}\wedge ...\wedge
dx_{i-1}\wedge Z_{i}[\varphi ,\psi ]dx_{i+1}\wedge ...\wedge dx_{m}
\label{1.4}
\end{equation}%
is an $(m-1)-$differential form on $\mathbb{R}^{m}.$

Consider now all such pairs $(\varphi (\lambda ),\psi (\mu ))\in \mathcal{H}%
_{0}^{\ast }\times \mathcal{H}_{0},$ $\lambda ,\mu \in \Sigma ,$ where $%
\Sigma \in \mathbb{C}^{p},$ $p\in \mathbb{Z}_{+},$ is some fixed measurable
space of parameters endowed with a bounded Lebesgue measure $\rho ,$ that
the differential form (\ref{1.4}) is exact, that is there exists such the
set of $(m-2)-$differential forms $\Omega ^{(m-2)}[\varphi (\lambda ),\psi
(\mu )]$ $\in \Lambda ^{m-2}(\mathbb{R}^{m};\mathbb{C}),$ $\lambda ,\mu \in
\Sigma ,$ on $\mathbb{R}^{m}$ satisfying the condition
\begin{equation}
Z^{(m-1)}[\varphi (\lambda ),\psi (\mu )]=d\Omega ^{(m-2)}[\varphi (\lambda
),\psi (\mu )].  \label{1.5}
\end{equation}%
Assume also that for any fixed element $\varphi (\lambda )\in \mathcal{H}%
_{0}^{\ast },$ $\lambda \in \Sigma ,$ \ the set $\mathcal{H}_{\varphi
}\subset \mathcal{H}_{0}$ of \ functions $\psi (\mu )\in \mathcal{H}_{0},$ $%
\mu \in \Sigma ,$ satisfying the condition (\ref{1.5}) is dense in $\mathcal{%
H},$ that is $\mathcal{\bar{H}}_{\varphi }=\mathcal{H}.$ Since the
relationship (\ref{1.5}) is semilinear in $(\varphi (\lambda ),\psi (\mu
))\in \mathcal{H}_{0}^{\ast }\times \mathcal{H}_{0},$ $\lambda ,\mu \in
\Sigma ,$ one gets easily that it holds for any pair $(\varphi (\lambda
),\psi (\mu ))\in \mathcal{H}_{0}^{\ast }\times \mathcal{H}_{0},$ $\lambda
,\mu \in \Sigma .$ Thus, taking into account that $d^{2}=0,$ one follows
from (\ref{1.3}) by integration over $\mathbb{R}^{m}$ that for any pair $%
(\varphi (\lambda ),\psi (\mu ))\in \mathcal{H}_{0}^{\ast }\times \mathcal{H}%
_{0}$ the identity $(<L^{\ast }\varphi (\lambda ),\psi (\mu )>)=(<\varphi
(\lambda ),L\psi (\mu )>),$ $\lambda ,\mu \in \Sigma ,$ holds, that is the
operator (\ref{1.1}) possesses its adjoint $L^{\ast }$ in $\mathcal{H}^{\ast
}.$ Another way to realize this condition is to take spaces $\mathcal{H}%
_{0}^{\ast }$ and $\ \mathcal{H}_{0}$ as solutions to the following linear
differential equations:%
\begin{eqnarray*}
\mathcal{H}_{0} &:&=\{\psi (\lambda )\in \mathcal{H}_{-}:L\psi (\lambda )=0,%
\text{ \ }\psi (\lambda )|_{x\in \Gamma }=0,\text{ }\lambda \in \Sigma \},%
\text{ } \\
\mathcal{H}_{0}^{\ast } &:&=\{\varphi (\lambda )\in \mathcal{H}_{-}^{\ast
}:L^{\ast }\varphi (\lambda )=0,\text{ \ }\varphi (\lambda )|_{x\in \Gamma
^{\ast }}=0,\text{ }\lambda \in \Sigma \},
\end{eqnarray*}%
where we have introduced following \cite{Be} a corresponding Hilbert-Schmidt
rigged chain of Hilbert spaces
\begin{equation}
\mathcal{H}_{+}\mathcal{\subset H\subset H}_{-},  \label{1.5a}
\end{equation}%
allowing to determine properly a set of generalized eigenfunctions of
extended operators $L$ $,L^{\ast }$ : $\mathcal{H}_{-}\rightarrow \mathcal{H}%
_{-},$ and $\ \Gamma ,\Gamma ^{\ast }$ $\subset \mathbb{R}^{m}$ are some
(n-1)-dimensional piece-wise smooth hypersurfaces imbedded into the
configuration space $\mathbb{R}^{m}.$ Let now $S(\sigma _{x}^{(m-2)},\sigma
_{x_{0}}^{(m-2)})$ denote an (m-1)-dimensional piece-wise smooth
hypersurface imbedded into $\mathbb{R}^{m}$ such that its boundaries $%
\partial S(\sigma _{x}^{(m-2)},\sigma _{x_{0}}^{(m-2)})=\sigma
_{x}^{(m-2)}-\sigma _{x_{0}}^{(m-2)},$ where $\sigma _{x}^{(m-2)}$ and $%
\sigma _{x_{0}}^{(m-2)}\in H_{m-2}(\mathbb{R}^{m};\mathbb{C})$ are some ($%
m-2)$-dimensional homological cycles from the homology group $H_{m-2}(%
\mathbb{R}^{m};\mathbb{C})$ of $\ $ $\mathbb{R}^{m},$ parametrized formally
by means of two points $x,x_{0}\in \mathbb{R}^{m}$ and related in some way
with the chosen above hypersurfaces $\Gamma $ \ and $\ \Gamma ^{\ast
}\subset \mathbb{R}^{m}.$ Then from (\ref{1.5}) based on the general Stokes
theorem \cite{Go,Te} one correspondingly gets easily that
\begin{equation*}
\int_{S(\sigma _{x}^{(m-2)},\sigma _{x_{0}}^{(m-2)})}Z^{(m-1)}[\varphi
(\lambda ),\psi (\mu )]=\int_{\partial S(\sigma _{x}^{(m-2)},\sigma
_{x_{0}}^{(m-2)})}\Omega ^{(m-2)}[\varphi (\lambda ),\psi (\mu )]=
\end{equation*}%
\begin{eqnarray}
&&\int_{\sigma _{x}^{(m-2)}}\Omega ^{(m-2)}[\varphi (\lambda ),\psi (\mu
)]-\int_{\sigma _{x_{0}}^{(m-2)}}\Omega ^{(m-2)}[\varphi (\lambda ),\psi
(\mu )]  \label{1.6} \\
&:&=\Omega _{x}[\varphi (\lambda ),\psi (\mu )]-\Omega _{x_{0}}[\varphi
(\lambda ),\psi (\mu )],  \notag
\end{eqnarray}%
\begin{equation*}
\int_{S(\sigma _{x}^{(m-2)},\sigma _{x_{0}}^{(m-2)})}\overset{\_}{Z}%
^{(m-1),\intercal }[\varphi (\lambda ),\psi (\mu )]=\int_{\partial S(\sigma
_{x}^{(m-2)},\sigma _{x_{0}}^{(m-2)})}\bar{\Omega}^{(m-2),\intercal
}[\varphi (\lambda ),\psi (\mu )]=
\end{equation*}%
\begin{eqnarray*}
&&\int_{\sigma _{x}^{(m-2)}}\bar{\Omega}^{(m-2),\intercal }[\varphi (\lambda
),\psi (\mu )]-\int_{\sigma _{x_{0}}^{(m-2)}}\bar{\Omega}^{(m-2),\intercal
}[\varphi (\lambda ),\psi (\mu )] \\
&:&=\Omega _{x}^{\ast }[\varphi (\lambda ),\psi (\mu )]-\Omega
_{x_{0}}^{\ast }[\varphi (\lambda ),\psi (\mu )]
\end{eqnarray*}%
for the set of functions $(\varphi (\lambda ),\psi (\mu ))\in \mathcal{H}%
^{\ast }\times \mathcal{H},$ $\lambda ,\mu \in \Sigma ,$ with kernels $%
\Omega _{x}[\varphi (\lambda ),\psi (\mu )],$ $\Omega _{x}^{\ast }[\varphi
(\lambda ),\psi (\mu )]\ $and $\ \Omega _{x_{0}}[\varphi (\lambda ),\psi
(\mu )],$ $\Omega _{x}^{\ast }[\varphi (\lambda ),\psi (\mu )],$ $\lambda
,\mu \in \Sigma ,$ acting naturally in the Hilbert space $L_{2}^{(\rho
)}(\Sigma ;\mathbb{C}),$ are assumed further to be nondegenerate in $%
L_{2}^{(\rho )}(\Sigma ;\mathbb{C})$ and satisfying the regularity conditions%
\begin{equation*}
\underset{x\rightarrow x_{0}}{lim}\Omega _{x}[\varphi (\lambda ),\psi (\mu
)]\ =\ \Omega _{x_{0}}[\varphi (\lambda ),\psi (\mu )],\text{ \ }\underset{%
x\rightarrow x_{0}}{lim}\Omega _{x}^{\ast }[\varphi (\lambda ),\psi (\mu )]\
=\ \Omega _{x_{0}}^{\ast }[\varphi (\lambda ),\psi (\mu )].
\end{equation*}%
Define now actions of the following two linear Delsarte permutations
operators $\mathbf{\Omega }:\mathcal{H}\rightarrow \mathcal{H}$ and $\mathbf{%
\Omega }^{\ast }:\mathcal{H}^{\ast }\rightarrow \mathcal{H}^{\ast }$ still
upon a fixed set of functions $(\varphi (\lambda ),\psi (\mu ))\in \mathcal{H%
}_{0}^{\ast }\times \mathcal{H}_{0},$ $\lambda ,\mu \in \Sigma ,:$%
\begin{equation*}
\tilde{\psi}(\lambda )=\mathbf{\Omega }(\psi (\lambda )):=\underset{\Sigma }{%
\int }d\rho (\eta )\underset{\Sigma }{\int }d\rho (\mu )\psi (\eta )\Omega
_{x}^{-1}[\varphi (\eta ),\psi (\mu )]\Omega _{x_{0}}[\varphi (\mu ),\psi
(\lambda )],
\end{equation*}%
\begin{equation}
\tilde{\varphi}(\lambda )=\mathbf{\Omega }^{\ast }(\varphi (\lambda )):=%
\underset{\Sigma }{\int }d\rho (\eta )\underset{\Sigma }{\int }d\rho (\mu
)\varphi (\eta )\Omega _{x}^{\ast ,-1}[\varphi (\eta ),\psi (\mu )]\Omega
_{x_{0}}^{\ast }[\varphi (\mu ),\psi (\lambda )].  \label{1.7}
\end{equation}%
Making use of the expressions (\ref{1.7}), based on arbitrariness of the
chosen \ set of functions $(\varphi (\lambda ),\psi (\mu ))\in \mathcal{H}%
_{0}^{\ast }\times \mathcal{H}_{0},$ $\lambda ,\mu \in \Sigma ,$ we can
easily retrieve the corresponding operator expressions for operators $%
\mathbf{\Omega }$ and $\mathbf{\Omega }^{\ast }:\mathcal{H}_{-}\mathcal{%
\rightarrow H}_{-},$ forcing the kernels $\Omega _{x_{0}}[\varphi (\lambda
),\psi (\mu )]$ \ and $\Omega _{x_{0}}^{\ast }[\varphi (\lambda ),\psi (\mu
)]),$ $\lambda ,\mu \in \Sigma ,$ to variate:%
\begin{eqnarray*}
\tilde{\psi}(\lambda ) &=&\underset{\Sigma }{\int }d\rho (\eta )\underset{%
\Sigma }{\int }d\rho (\mu )\underset{\Sigma }{\int }d\rho (\nu )\psi (\eta
)\Omega _{x}[\varphi (\eta ),\psi (\mu )]\Omega _{x}^{-1}[\varphi (\mu
),\psi (\lambda )] \\
&&-\underset{\Sigma }{\int }d\rho (\eta )\underset{\Sigma }{\int }d\rho (\mu
)\underset{\Sigma }{\int }d\rho (\nu )\underset{\Sigma }{\int }d\rho (\xi
)\psi (\eta )\Omega _{x}^{-1}[\varphi (\eta ),\psi (\mu )]\times \\
&&\times \int_{S(\sigma _{x}^{(m-2)},\sigma
_{x_{0}}^{(m-2)})}Z^{(m-1)}[\varphi (\mu ),\psi (\lambda )])
\end{eqnarray*}%
\begin{eqnarray*}
&=&\psi (\lambda )-\underset{\Sigma }{\int }d\rho (\eta )\underset{\Sigma }{%
\int }d\rho (\mu )\underset{\Sigma }{\int }d\rho (\nu )\underset{\Sigma }{%
\int }d\rho (\xi )\psi (\eta )\Omega _{x}^{-1}[\varphi (\eta ),\psi (\mu
)]\times \\
&&\times \Omega _{x_{0}}[\varphi (\mu ),\psi (\nu )]\Omega
_{x_{0}}^{-1}[\varphi (\nu ),\psi (\xi )]\int_{S(\sigma _{x}^{(m-2)},\sigma
_{x_{0}}^{(m-2)})}Z^{(m-1)}[\varphi (\xi ),\psi (\mu )]
\end{eqnarray*}%
\begin{equation*}
=\psi (\lambda )-\underset{\Sigma }{\int }d\rho (\eta )\underset{\Sigma }{%
\int }d\rho (\mu )\tilde{\psi}\Omega _{x_{0}}^{-1}[\varphi (\eta ),\psi (\mu
)]\int_{S(\sigma _{x}^{(m-2)},\sigma _{x_{0}}^{(m-2)})}Z^{(m-1)}[\varphi
(\mu ),\psi (\lambda )]
\end{equation*}%
\begin{equation*}
=(\mathbf{1}-\underset{\Sigma }{\int }d\rho (\eta )\underset{\Sigma }{\int }%
d\rho (\mu )\tilde{\psi}(\eta )\Omega _{x_{0}}^{-1}[\varphi (\eta ),\psi
(\mu )]\times
\end{equation*}%
\begin{equation*}
\times \int_{S(\sigma _{x}^{(m-2)},\sigma
_{x_{0}}^{(m-2)})}Z^{(m-1)}[\varphi (\mu ),\cdot ])\text{ }\psi (\lambda ):=%
\mathbf{\Omega }\cdot \psi (\lambda );
\end{equation*}%
\begin{eqnarray*}
\tilde{\varphi}(\lambda ) &=&\underset{\Sigma }{\int }d\rho (\eta )\underset{%
\Sigma }{\int }d\rho (\mu )\varphi (\eta )\Omega _{x}^{-1\ast }[\varphi (\mu
),\psi (\eta )]\Omega _{x}^{\ast }[\varphi (\lambda ),\psi (\mu )] \\
&&-\underset{\Sigma }{\int }d\rho (\eta )\underset{\Sigma }{\int }d\rho (\mu
)\varphi (\eta )\Omega _{x_{0}}^{\ast ,-1}[\varphi (\mu ),\psi (\eta
)]\int_{S(\sigma _{x}^{(m-2)},\sigma _{x_{0}}^{(m-2)})}\bar{Z}%
^{(m-1),\intercal }[\varphi (\lambda ),\psi (\mu )]
\end{eqnarray*}%
\begin{eqnarray*}
&=&\varphi (\lambda )-\underset{\Sigma }{\int }d\rho (\eta )\underset{\Sigma
}{\int }d\rho (\xi )\underset{\Sigma }{\int }d\rho (\mu )\underset{\Sigma }{%
\int }d\rho (\nu )\varphi (\eta )\Omega _{x}^{\ast ,-1}[\varphi (\nu ),\psi
(\eta )]\times \\
&&\times \Omega _{x_{_{0}}}^{\ast }[\varphi (\xi ),\psi (\nu )]\Omega
_{x_{0}}^{\ast ,-1}[\varphi (\mu ),\psi (\xi )]\int_{S(\sigma
_{x}^{(m-2)},\sigma _{x_{0}}^{(m-2)})}\bar{Z}^{(m-1),\intercal }[\varphi
(\lambda ),\psi (\mu )]
\end{eqnarray*}%
\begin{equation}
=(\mathbf{1}-\underset{\Sigma }{\int }d\rho (\eta )\underset{\Sigma }{\int }%
d\rho (\mu )\tilde{\varphi}(\eta )\Omega _{x_{_{0}}}^{\ast ,-1}[\varphi (\mu
),\psi (\eta )]\times  \label{1.8}
\end{equation}%
\begin{equation*}
\times \int_{S(\sigma _{x}^{(m-2)},\sigma _{x_{0}}^{(m-2)})}\bar{Z}%
^{(m-1),\intercal }[\cdot ,\psi (\mu )])\text{ }\varphi (\lambda ):=\mathbf{%
\Omega }^{\ast }\cdot \varphi (\lambda ),
\end{equation*}%
where, by definition,
\begin{equation*}
\mathbf{\Omega }:=\mathbf{1}-\underset{\Sigma }{\int }d\rho (\eta )\underset{%
\Sigma }{\int }d\rho (\mu )\tilde{\psi}(\eta )\Omega _{x_{0}}^{-1}[\varphi
(\eta ),\psi (\mu )]\int_{S(\sigma _{x}^{(m-2)},\sigma
_{x_{0}}^{(m-2)})}Z^{(m-1)}[\varphi (\mu ),\cdot ]
\end{equation*}%
\begin{equation}
\mathbf{\Omega }^{\ast }:=\mathbf{1}-\underset{\Sigma }{\int }d\rho (\eta )%
\underset{\Sigma }{\int }d\rho (\mu )\tilde{\varphi}(\eta )\Omega
_{x_{_{0}}}^{\ast ,-1}[\varphi (\mu ),\psi (\eta )]\int_{S(\sigma
_{x}^{(m-2)},\sigma _{x_{0}}^{(m-2)})}\bar{Z}^{(m-1),\intercal }[\cdot ,\psi
(\mu )]  \label{1.9}
\end{equation}%
are of Volterra type multidimensional integral operators. It is to be noted
here that now elements $(\varphi (\lambda ),\psi (\mu ))\in \mathcal{H}%
_{0}^{\ast }\times \mathcal{H}_{0}$ and $(\tilde{\varphi}(\lambda ),\tilde{%
\psi}(\mu ))\in \mathcal{\tilde{H}}_{0}^{\ast }\times \mathcal{\tilde{H}}%
_{0},$ $\lambda ,\mu \in \Sigma ,$ inside the operator expressions (\ref{1.9}%
) are arbitrary but fixed. Therefore, the operators (\ref{1.9}) realize an
extension of their actions (\ref{1.7}) on a fixed pair of functions $%
(\varphi (\lambda ),\psi (\mu ))\in \mathcal{H}_{0}^{\ast }\times \mathcal{H}%
_{0},$ $\lambda ,\mu \in \Sigma ,$ upon the whole functional space $\mathcal{%
H}^{\ast }\times \mathcal{H}.$

Due to the symmetry of expressions (\ref{1.7}) and (\ref{1.9}) with respect
to two sets of functions $(\varphi (\lambda ),\psi (\mu ))\in \mathcal{H}%
_{0}^{\ast }\times \mathcal{H}_{0}$ and $(\tilde{\varphi}(\lambda ),\tilde{%
\psi}(\mu ))\in \mathcal{\tilde{H}}_{0}^{\ast }\times \mathcal{\tilde{H}}%
_{0},$ $\lambda ,\mu \in \Sigma ,$ it is very easy to state the following
lemma.

\begin{lemma}
Operators (\ref{1.9}) are bounded and invertible of Volterra type
expressions in $\mathcal{H}^{\ast }\times \mathcal{H}$ \ whose inverse are
given as follows:%
\begin{equation}
\mathbf{\Omega }^{-1}:=\mathbf{1}-\underset{\Sigma }{\int }d\rho (\eta )%
\underset{\Sigma }{\int }d\rho (\mu )\psi (\eta )\tilde{\Omega}_{x_{0}}^{-1}[%
\tilde{\varphi}(\eta ),\tilde{\psi}(\mu )]\int_{S(\sigma _{x}^{(m-2)},\sigma
_{x_{0}}^{(m-2)})}Z^{(m-1)}[\tilde{\varphi}(\mu ),\cdot ]  \label{1.10}
\end{equation}%
\begin{equation*}
\mathbf{\Omega }^{\ast ,-1}:=\mathbf{1}-\underset{\Sigma }{\int }d\rho (\eta
)\underset{\Sigma }{\int }d\rho (\mu )\varphi (\eta )\Omega
_{x_{_{0}}}^{\ast ,-1}[\tilde{\varphi}(\mu ),\tilde{\psi}(\eta
)]\int_{S(\sigma _{x}^{(m-2)},\sigma _{x_{0}}^{(m-2)})}\bar{Z}%
^{(m-1),\intercal }[\cdot ,\tilde{\psi}(\mu )]
\end{equation*}%
where two sets of functions $(\tilde{\varphi}(\lambda ),\tilde{\psi}(\mu
))\in \mathcal{H}_{0}^{\ast }\times \mathcal{H}_{0}$ and $(\tilde{\varphi}%
(\lambda ),\tilde{\psi}(\mu ))\in \mathcal{\tilde{H}}_{0}^{\ast }\times
\mathcal{\tilde{H}}_{0},$ $\lambda ,\mu \in \Sigma ,$ are taken arbitrary
but fixed.
\end{lemma}

For the expressions (\ref{1.10}) to be compatible with mappings (\ref{1.7})
the following actions must hold:%
\begin{equation*}
\psi (\lambda )=\mathbf{\Omega }^{-1}\cdot \tilde{\psi}(\lambda )=\underset{%
\Sigma }{\int }d\rho (\eta )\underset{\Sigma }{\int }d\rho (\mu )\tilde{\psi}%
(\eta )\tilde{\Omega}_{x}^{-1}[\tilde{\varphi}(\eta ),\tilde{\psi}(\mu )]%
\tilde{\Omega}_{x_{0}}[\tilde{\varphi}(\mu ),\tilde{\psi}(\lambda )]),
\end{equation*}%
\begin{equation}
\varphi (\lambda )=\mathbf{\Omega }^{\ast ,-1}\cdot \tilde{\varphi}(\lambda
)=\underset{\Sigma }{\int }d\rho (\eta )\underset{\Sigma }{\int }d\rho (\mu )%
\tilde{\varphi}(\eta )\tilde{\Omega}_{x}^{\ast ,-1}[\tilde{\varphi}(\mu ),%
\tilde{\psi}(\eta )]\tilde{\Omega}_{x_{0}}^{\ast }[\tilde{\varphi}(\lambda ),%
\tilde{\psi}(\mu )]),  \label{1.11}
\end{equation}%
where for any two sets of functions $(\tilde{\varphi}(\lambda ),\tilde{\psi}%
(\mu ))\in \mathcal{H}_{0}^{\ast }\times \mathcal{H}_{0}$ and $(\tilde{%
\varphi}(\lambda ),\tilde{\psi}(\mu ))\in \mathcal{\tilde{H}}_{0}^{\ast
}\times \mathcal{\tilde{H}}_{0},$ $\lambda ,\mu \in \Sigma ,$ the next
relationship is satisfied:%
\begin{equation*}
(<\tilde{L}^{\ast }\tilde{\varphi}(\lambda ),\tilde{\psi}(\mu )>-<\tilde{%
\varphi}(\lambda ),\tilde{L}\tilde{\psi}(\mu )>)dx=d(\tilde{Z}^{(m-1)}[%
\tilde{\varphi}(\lambda ),\tilde{\psi}(\mu )]),
\end{equation*}%
\begin{equation*}
\tilde{Z}^{(m-1)}[\tilde{\varphi}(\lambda ),\tilde{\psi}(\mu )]=d\tilde{%
\Omega}^{(m-2)}[\tilde{\varphi}(\lambda ),\tilde{\psi}(\mu )].
\end{equation*}%
\begin{equation}
\tilde{L}:=\mathbf{\Omega }L\mathbf{\Omega }^{-1},\text{ \ }\tilde{L}^{\ast
}:=\mathbf{\Omega }^{\ast }L^{\ast }\mathbf{\Omega }^{\ast ,-1},\text{ }
\label{1.12}
\end{equation}%
Moreover, the expressions $\tilde{L}:\mathcal{H}\rightarrow \mathcal{H}$ and
$\tilde{L}^{\ast }:\mathcal{H}^{\ast }\rightarrow \mathcal{H}^{\ast }$ must
in the result be differential too. Since this condition determines properly
Delsarte transmutation operators (\ref{1.10}), we need to state the
following theorem.

\begin{theorem}
The pair of operator expressions $\tilde{L}:=\mathbf{\Omega }L\mathbf{\Omega
}^{-1}$ and $\tilde{L}^{\ast }:=\mathbf{\Omega }^{\ast }L^{\ast }\mathbf{%
\Omega }^{\ast ,-1}$ is purely differential on the whole space $\mathcal{H}%
^{\ast }\times \mathcal{H}$ \ for any suitably chosen hyper-surface $%
S(\sigma _{x}^{(m-2)},\sigma _{x_{0}}^{(m-2)})\subset \mathbb{R}^{m}.$
\end{theorem}

\begin{proof}
For proving the theorem it is necessary to show that the formal
pseudo-differential expressions corresponding to operators $\tilde{L}$ and $%
\tilde{L}^{\ast }$ contain no integral elements. Making use of an idea
devised in \cite{SP,Ni}, one can formulate such a lemma.
\end{proof}

\begin{lemma}
A pseudo-differential operator $L:\mathcal{H}\rightarrow \mathcal{H}$ is
purely differential iff the following equality \
\begin{equation}
(<h,(L\frac{\partial ^{|\alpha |}}{\partial x^{\alpha }})_{+}f>)=(<h,L_{+}%
\frac{\partial ^{|\alpha |}}{\partial x^{\alpha }}f>)  \label{1.13}
\end{equation}%
holds for any $|\alpha |\in \mathbb{Z}_{+}$ and all $(h,f)\in \mathcal{H}%
^{\ast }\times \mathcal{H},$ that is the condition (\ref{1.13}) is
equivalent to the equality $L_{+}=L,$ where, as usually, the sign "$%
(...)_{+}"$ \ means the purely differential part of the corresponding
expression inside the bracket.
\end{lemma}

Based now on this Lemma and exact expressions of operators (\ref{1.9}),
similarly to calculations done in \cite{SP}, one shows right away that
operators $\tilde{L}$ and $\tilde{L}^{\ast },$ depending correspondingly
only both on the homological cycles $\sigma _{x}^{(m-2)},\sigma
_{x_{0}}^{(m-2)}\in H_{m-2}(\mathbb{R}^{m};\mathbb{C}),$ marked by points $%
x,x_{0}\in \mathbb{R}^{m},$ and on two sets of functions $(\tilde{\varphi}%
(\lambda ),\tilde{\psi}(\mu ))\in \mathcal{H}_{0}^{\ast }\times \mathcal{H}%
_{0}$ and $(\tilde{\varphi}(\lambda ),\tilde{\psi}(\mu ))\in \mathcal{\tilde{%
H}}_{0}^{\ast }\times \mathcal{\tilde{H}}_{0},$ $\lambda ,\mu \in \Sigma ,$
are purely differential, thereby finishing the proof.$\blacktriangleright $

\section{\protect\bigskip The general differential-geometric and topological
structure of Delsarte transmutation operators}

\setcounter{equation}{0}Let $M:=\mathbb{\bar{R}}^{m}$ denote a suitably
compactified metric space of dimension $m=dimM\in \mathbb{Z}_{+}$ (without
boundary) and define some finite set $\mathcal{L}$ \ of smooth commuting to
each other linear differential operators
\begin{equation}
L_{j}(x;\partial ):=\sum_{|\alpha |=0}^{n(L_{j})}a_{\alpha
}^{(j)}(x)\partial ^{|\alpha |}/\partial x^{\alpha },  \label{3.1}
\end{equation}%
$x\in M,$ with Schwatrz coefficients $a_{\alpha }^{(j)}\in \mathcal{S}(M;End%
\mathbb{C}^{N}),$ $|\alpha |=\overline{0,n(L_{j})},$ $n(L_{j})\in \mathbb{Z}%
_{+},$ $j=\overline{1,m},$ and acting in the Hilbert space $\mathcal{H}%
:=L_{2}(M;\mathbb{C}^{N}).$ It is assumed that domains $D(L_{j}):=D(\mathcal{%
L})\subset \mathcal{H},$ $j=\overline{1,m},$ are dense in $\mathcal{H}.$

Consider now a generalized external differentiation operator $d_{\mathcal{L}%
} $ :$\Lambda (M;\mathcal{H)\rightarrow }\Lambda (M;\mathcal{H)}$ acting in
the Grassmann algebra $\Lambda (M;\mathcal{H)}$ as follows: for any $\beta
^{(k)}\in \Lambda ^{k}(M;\mathcal{H)},$ $k=\overline{0,m},$
\begin{equation}
d_{\mathcal{L}}\beta ^{(k)}:=\sum_{j=1}^{m}dx_{j}\wedge L_{j}(x;\partial
)\beta ^{(k)}\in \Lambda ^{k+1}(M;\mathcal{H)}.  \label{3.2}
\end{equation}%
It is easy to see that the operation (\ref{3.2}) in the case $%
L_{j}(x;\partial ):=\partial /\partial x_{j},$ $j=\overline{1,m},$ coincides
exactly with the standard external differentiation $d=\sum_{j=1}^{m}dx_{j}%
\wedge \partial /\partial x_{j}$ on the Grassmann algebra $\Lambda (M;%
\mathcal{H)}.$ Making use of the operation (\ref{3.2}) on $\Lambda (M;%
\mathcal{H)},$ one can construct the following generalized de Rham complex

\begin{equation}
\mathcal{H}\rightarrow \Lambda ^{0}(M;\mathcal{H)}\overset{d_{\mathcal{L}}}{%
\rightarrow }\Lambda ^{1}(M;\mathcal{H)}\overset{d_{\mathcal{L}}}{%
\rightarrow }...\overset{d_{\mathcal{L}}}{\rightarrow }\Lambda ^{m}(M;%
\mathcal{H)}\overset{d_{\mathcal{L}}}{\rightarrow }0.  \label{3.3}
\end{equation}%
The following important property concerning the complex (\ref{3.3}) holds.

\begin{lemma}
The co-chain complex (\ref{3.3}) is exact.
\end{lemma}

\begin{proof}
It follows easily from the equality $d_{\mathcal{L}}d_{\mathcal{L}}=0$
holding due to the commutation of operators (\ref{3.1}) .$\triangleright $
\end{proof}

\bigskip Below we will follow the ideas developed before in \cite{Sk}. A
differential form $\beta \in \Lambda (M;\mathcal{H)}$ will be called $d_{%
\mathcal{L}}$-closed if $d_{\mathcal{L}}\beta =0,$ and a form $\gamma \in
\Lambda (M;\mathcal{H)}$ will be called $d_{\mathcal{L}}$-homological to
zero if there exists on $M$ such a form $\omega \in \Lambda (M;\mathcal{H)}$
that $\gamma =d_{\mathcal{L}}\omega .$

Consider now the standard algebraic Hodge star-operation
\begin{equation}
\star :\Lambda ^{k}(M;\mathcal{H)\rightarrow }\Lambda ^{m-k}(M;\mathcal{H)},
\label{3.3a}
\end{equation}%
$k=\overline{0,m},$ as follows \cite{Ch}: if $\beta \in \Lambda ^{k}(M;%
\mathcal{H)},$ then the form $\star \beta \in \Lambda ^{m-k}(M;\mathcal{H)}$
is such that:

i) $\ (m-k)$-dimensional volume $|\star \beta |$ of the form $\star \beta $
equals $k$-dimensional volume $|\beta |$ of the form $\beta ;$

ii) the $m$-dimensional measure $\bar{\beta}^{\intercal }\wedge \star \beta $
$>0$ under the fixed orientation on $M.$

Define also on the space $\Lambda (M;\mathcal{H)}$ the following natural
scalar product: for any $\beta ,\gamma \in \Lambda ^{k}(M;\mathcal{H)},$ $k=%
\overline{0,m},$%
\begin{equation}
(<\beta ,\gamma >):=\int_{M}\bar{\beta}^{\intercal }\wedge \star \gamma .
\label{3.4}
\end{equation}%
Subject to the scalar product (\ref{3.4}) we can naturally construct the
corresponding Hilbert space
\begin{equation*}
\mathcal{H}_{\Lambda }(M):=\overset{m}{\underset{k=0}{\oplus }}\mathcal{H}%
_{\Lambda }^{k}(M)
\end{equation*}
well suitable for our further consideration. Notice also here that the Hodge
star $\star $-operation satisfies the following easily checkable property:
for any $\beta ,\gamma \in \mathcal{H}_{\Lambda }^{k}(M),$ $k=\overline{0,m}%
, $%
\begin{equation}
(<\beta ,\gamma >)=(<\star \beta ,\star \gamma >),  \label{3.5}
\end{equation}%
that is the Hodge operation $\star :\mathcal{H}_{\Lambda }(M)\mathcal{%
\rightarrow H}_{\Lambda }(M)$ is isometry and its standard adjoint with
respect to the scalar product (\ref{3.4}) operation $(\star )^{^{\prime
}}=(\star )^{-1}.$

Denote by $d_{\mathcal{L}}^{\prime }$ the formally adjoint expression to the
external weak differential operation $d_{\mathcal{L}}$ :$\mathcal{H}%
_{\Lambda }(M)\mathcal{\rightarrow H}_{\Lambda }(M)$ in the Hilbert space $%
\mathcal{H}_{\Lambda }(M).$ Making now use of the operations $d_{\mathcal{L}%
}^{\prime }$ and $d_{\mathcal{L}}$ in $\mathcal{H}_{\Lambda }(M)$ one can
naturally define \cite{Ch} the generalized Laplace-Hodge operator $\Delta _{%
\mathcal{L}}:\mathcal{H}_{\Lambda }(M)\rightarrow \mathcal{H}_{\Lambda }(M)$
as
\begin{equation}
\Delta _{\mathcal{L}}:=d_{\mathcal{L}}^{\prime }d_{\mathcal{L}}+d_{\mathcal{L%
}}d_{\mathcal{L}}^{\prime }.  \label{3.6}
\end{equation}%
Take a form $\beta \in \mathcal{H}_{\Lambda }(M)$ satisfying the equality
\begin{equation}
\Delta _{\mathcal{L}}\beta =0.  \label{3.7}
\end{equation}%
Such a form is called \textit{harmonic}. One can also verify that a harmonic
form $\beta \in \mathcal{H}_{\Lambda }(M)$ satisfies simultaneously the
following two adjoint conditions:%
\begin{equation}
d_{\mathcal{L}}^{\prime }\beta =0,\text{ \ \ }d_{\mathcal{L}}\beta =0,
\label{3.8}
\end{equation}%
easily stemming from (\ref{3.6}) and (\ref{3.8}).

\bigskip\ It is not hard to check that the following differential operation
in $\mathcal{H}_{\Lambda }(M)$%
\begin{equation}
d_{\mathcal{L}}^{\ast }:=\star d_{\mathcal{L}}^{\prime }(\star )^{-1}
\label{3.9}
\end{equation}%
defines the usual \cite{Go,Te} external anti-differential operation in $%
\mathcal{H}_{\Lambda }(M).$ The corresponding dual to (\ref{3.3}) complex
\begin{equation}
\mathcal{H}\rightarrow \Lambda ^{0}(M;\mathcal{H)}\overset{d_{\mathcal{L}%
}^{\ast }}{\rightarrow }\Lambda ^{1}(M;\mathcal{H)}\overset{d_{\mathcal{L}%
}^{\ast }}{\rightarrow }...\overset{d_{\mathcal{L}}^{\ast }}{\rightarrow }%
\Lambda ^{m}(M;\mathcal{H)}\overset{d_{\mathcal{L}}^{\ast }}{\rightarrow }0
\label{3.9a}
\end{equation}%
is evidently exact too, as the property $d_{\mathcal{L}}^{\ast }d_{\mathcal{L%
}}^{\ast }=0$ holds due to the definition (\ref{3.6}).

\bigskip Denote further by $\mathcal{H}_{\Lambda (\mathcal{L})}^{k}(M),$ $k=%
\overline{0,m},$ the co$\hom $o$\log $y groups of $d_{\mathcal{L}}$-closed
and by $\mathcal{H}_{\Lambda (\mathcal{L}^{\ast })}^{k}(M),$ $k=\overline{0,m%
},$ the co$\hom $o$\log $y groups of $d_{\mathcal{L}}^{\ast }$-closed
differential forms, correspondingly, and by $\mathcal{H}_{\Lambda (\mathcal{L%
}^{\ast }\mathcal{L})}^{k}(M),$ $k=\overline{0,m},$ the abelian groups of
harmonic differential forms from the Hilbert sub-spaces $\mathcal{H}%
_{\Lambda }^{k}(M),$ $k=\overline{0,m}.$ Before formulating next results,
define the standard Hilbert-Schmidt rigged chain \cite{Be} of positive and
negative Hilbert spaces \ of differential forms
\begin{equation}
\mathcal{H}_{\Lambda ,+}^{k}(M)\subset \mathcal{H}_{\Lambda }^{k}(M)\subset
\mathcal{H}_{\Lambda ,-}^{k}(M)  \label{3.9b}
\end{equation}%
and the corresponding rigged chains of Hilbert sub-spaces for harmonic%
\begin{equation}
\mathcal{H}_{\Lambda (\mathcal{L}^{\ast }\mathcal{L}),+}^{k}(M)\subset
\mathcal{H}_{\Lambda (\mathcal{L}^{\ast }\mathcal{L})}^{k}(M)\subset
\mathcal{H}_{\Lambda (\mathcal{L}^{\ast }\mathcal{L}),-}^{k}(M),
\label{3.9c}
\end{equation}%
and cohomology groups:%
\begin{eqnarray}
\mathcal{H}_{\Lambda (\mathcal{L}),+}^{k}(M) &\subset &\mathcal{H}_{\Lambda (%
\mathcal{L})}^{k}(M)\subset \mathcal{H}_{\Lambda (\mathcal{L}),-}^{k}(M),
\label{3.9d} \\
\mathcal{H}_{\Lambda (\mathcal{L}^{\ast }),+}^{k}(M) &\subset &\mathcal{H}%
_{\Lambda (\mathcal{L}^{\ast })}^{k}(M)\subset \mathcal{H}_{\Lambda (%
\mathcal{L}^{\ast }),-}^{k}(M),  \notag
\end{eqnarray}%
for any $k=\overline{0,m}.$ Now by reasonings similar to those in \cite%
{Ch,Te} one can formulate the following a little generalized de Rham-Hodge
theorem.

\begin{theorem}
The groups of harmonic forms $\mathcal{H}_{\Lambda (\mathcal{L}^{\ast }%
\mathcal{L}),-}^{k}(M),$ $k=\overline{0,m},$ are, correspondingly,
isomorphic to the cohomology groups $(H^{k}(M;\mathbb{C}))^{\Sigma },$ $k=%
\overline{0,m},$ where $H^{k}(M;\mathbb{C)}$ is the $k-$th cohomology group
of the manifold $M$ with complex coefficients, $\Sigma \subset $ $\mathbb{C}%
^{p}$ is a set of suitable "spectral" parameters marking the linear space of
independent $d_{\mathcal{L}}^{\ast }$-closed 0-forms from $\mathcal{H}%
_{\Lambda (\mathcal{L}),-}^{0}(M)$ and, moreover, the following direct sum
decompositions
\begin{equation}
\mathcal{H}_{\Lambda (\mathcal{L}^{\ast }\mathcal{L}),-}^{k}(M)\oplus \Delta
\mathcal{H}_{-}^{k}(M)=\mathcal{H}_{\Lambda ,-}^{k}(M)=\mathcal{H}_{\Lambda (%
\mathcal{L}^{\ast }\mathcal{L}),-}^{k}(M)\oplus d_{\mathcal{L}}\mathcal{H}%
_{\Lambda ,-}^{k-1}(M)\oplus d_{\mathcal{L}}^{^{\prime }}\mathcal{H}%
_{\Lambda ,-}^{k+1}(M)  \label{3.9e}
\end{equation}%
hold for any $k=\overline{0,m}.$
\end{theorem}

Another variant of the statement similar to that above was formulated in
\cite{Sk} and reads as the following generalized de Rham-Hodge-Skrypnik
theorem.

\begin{theorem}
(See Skrypnik I.V. \cite{Sk} \ The generalized cohomology groups $\mathcal{H}%
_{\Lambda (\mathcal{L}),-}^{k}(M),k=\overline{0,m},$ are isomorphic,
correspondingly, to the cohomology groups $(H^{k}(M;\mathbb{C}))^{\Sigma },$
$k=\overline{0,m}.$
\end{theorem}

A proof of this theorem is based on some special sequence \cite{Sk} of
differential Lagrange type identities. Define the following closed subspace
\begin{equation}
\mathcal{H}_{0}^{\ast }:=\{\varphi (\lambda )\in \mathcal{H}_{\Lambda (%
\mathcal{L}^{\ast }),-}^{0}(M):d_{\mathcal{L}}^{\ast }\varphi (\lambda )=0,%
\text{ }\varphi (\lambda )|_{\Gamma ^{\ast }}=0,\text{ }\lambda \in \Sigma \}
\label{3.10}
\end{equation}%
for some smooth $(m-1)$-dimensional hypersurface $\Gamma ^{\ast }\subset M$
and $\Sigma \subset (\sigma (\mathcal{\tilde{L}})\cap \bar{\sigma}(\mathcal{L%
}^{\ast }))\times \Sigma _{\sigma }\subset \mathbb{C}^{p},$ where $\mathcal{H%
}_{\Lambda (\mathcal{L}^{\ast }),-}^{0}(M)$ is, as above, a suitable
Hilbert-Schmidt rigged \cite{Be} zero-order cohomology group Hilbert space
from the chain given by (\ref{3.9e}), $\sigma (\mathcal{\tilde{L}})$ and $%
\sigma (\mathcal{L}^{\ast })$ are , correspondingly, mutual spectra of the
sets of operators $\mathcal{\tilde{L}}$ and $\mathcal{L}^{\ast }.$ Thereby
the dimension $\dim $ $\mathcal{H}_{0}^{\ast }=card$ $\Sigma $ is assumed to
be known.

The next lemma stated by Skrypnik I.V. \cite{Sk} being fundamental for the
proof holds.

\begin{lemma}
(See Skrypnik\ I.V. \cite{Sk} ) There exists a set of differential $k$-forms
$Z^{(k+1)}[\varphi (\lambda ),d_{\mathcal{L}}\psi ]\in \Lambda ^{k+1}(M;%
\mathcal{H}),$ $k=\overline{0,m},$ and a set of $k$-forms $Z^{(k)}[\varphi
(\lambda ),\psi ]\in \Lambda ^{k}(M;\mathcal{H}),$ $k=\overline{0,m},$
parametrized by a set $\Sigma \ni \lambda $ and semilinear in $(\varphi
(\lambda ),\psi )\in \mathcal{H}_{0}^{\ast }\times \mathcal{H}_{\Lambda
,-}^{k}(M),$ such that%
\begin{equation}
Z^{(k+1)}[\varphi (\lambda ),d_{\mathcal{L}}\psi ]=dZ^{(k)}[\varphi (\lambda
),\psi ]  \label{3.11}
\end{equation}%
for all $k=\overline{0,m}$ \ and $\lambda \in \Sigma .$
\end{lemma}

\begin{proof}
A proof is based on the following generalized Lagrange type identity holding
for any pair $(\varphi (\lambda ),\psi )\in \mathcal{H}_{0}^{\ast }\times
\mathcal{H}_{\Lambda ,-}^{k}(M):$%
\begin{eqnarray}
0 &=&<d_{\mathcal{L}}^{\ast }\varphi (\lambda ),\star (\psi \wedge \bar{%
\gamma})>:=<\star d_{\mathcal{L}}^{\prime }(\star )^{-1}\varphi (\lambda
),\star (\psi \wedge \bar{\gamma})>  \label{3.12} \\
&=&<d_{\mathcal{L}}^{\prime }(\star )^{-1}\varphi (\lambda ),\psi \wedge
\bar{\gamma}>=<(\star )^{-1}\varphi (\lambda ),d_{\mathcal{L}}\psi \wedge
\bar{\gamma}>  \notag
\end{eqnarray}%
\begin{equation*}
+Z^{(k+1)}[\varphi (\lambda ),d_{\mathcal{L}}\psi ]\wedge \bar{\gamma}\text{
}\equiv <(\star )^{-1}\varphi (\lambda ),d_{\mathcal{L}}\psi \wedge \bar{%
\gamma}>+dZ^{(k)}[\varphi (\lambda ),\psi ]\wedge \bar{\gamma}
\end{equation*}%
where $Z^{(k+1)}[\varphi (\lambda ),d_{\mathcal{L}}\psi ]\in \Lambda
^{k+1}(M;\mathbb{C}),$ $k=\overline{0,m},$ and $Z^{(k)}[\varphi (\lambda
),\psi ]\in \Lambda ^{k-1}(M;\mathbb{C}),$ $k=\overline{0,m},$ are some
semilinear differential forms parametrized by a parameter $\lambda \in
\Sigma ,$ and $\bar{\gamma}\in \Lambda ^{m-k-1}(M;\mathbb{C})$ is arbitrary
constant $(m-k-1)$-form. Thereby, the semilinear differential $k$-forms $%
Z^{(k+1)}[\varphi (\lambda ),d_{\mathcal{L}}\psi ]\in \Lambda ^{k+1}(M;%
\mathbb{C}),$ $k=\overline{0,m},$ and $k$-forms $Z^{(k)}[\varphi (\lambda
),\psi ]\in \Lambda ^{k}(M;\mathbb{C}),$ $k=\overline{0,m},$ $\lambda \in
\Sigma ,$ constructed above exactly constitute those searched for in the
Lemma.$\triangleright $
\end{proof}

Based now on this Lemma 3.3 one can construct the cohomology group
isomorphism claimed in the Theorem 3.2 formulated above. Namely, following
\cite{Sk}, let us take some simplicial partition $K(M)$ of the manifold $M$
and introduce linear mappings $B_{\lambda }^{(k)}:\mathcal{H}_{\Lambda
,-}^{k}(M)\rightarrow C_{k}(M),$ $k=\overline{0,m},$ $\lambda \in \Sigma ,$
where $C_{k}(M),$ $k=\overline{0,m},$ are the free abelian groups over the
field $\mathbb{C}$ generated, correspondingly, by all $k$-chains of
simplexes $S^{(k)}\in C_{k}(M),$ $k=\overline{0,m},$ of the simplicial \cite%
{Te} complex $K(M)$ as follows:%
\begin{equation}
B_{\lambda }^{(k)}(\psi ):=\sum_{S^{(k)}\in
C_{k}(M)}S^{(k)}\int_{S^{(k)}}Z^{(k)}[\varphi (\lambda ),\psi ]  \label{3.13}
\end{equation}%
with $\psi \in \mathcal{H}_{\Lambda }^{k}(M),$ $k=\overline{0,m}.$ The
following theorem based on mappings (\ref{3.13}) holds.

\begin{theorem}
(See Skrypnik I.V. \cite{Sk} ) The set of operations (\ref{3.13})
parametrized by $\lambda \in \Sigma $ realizes the cohomology groups
isomorphism formulated in the Theorem 3.2.
\end{theorem}

\begin{proof}
A proof of this theorem one can get passing over in (\ref{3.13}) to the
corresponding cohomology $\mathcal{H}_{\Lambda (\mathcal{L}),-}^{k}(M)$ and
homology $H_{k}(M;\mathbb{C})$ groups of $M$ $\ $for every $k=\overline{0,m}%
. $ If one to take an element $\psi :=\psi (\mu )\in \mathcal{H}_{\Lambda (%
\mathcal{L}),-}^{k}(M),$ $k=\overline{0,m},$ solving the equation $d_{%
\mathcal{L}}\psi (\mu )=0$ with $\mu \in \Sigma _{k}$ being some set of \
the related "spectral" parameters marking elements of the subspace $\mathcal{%
H}_{\Lambda (\mathcal{L}),-}^{k}(M),$ then one finds easily from (\ref{3.13}%
) and the identity (\ref{3.12}) that
\begin{equation}
dZ^{(k)}[\varphi (\lambda ),\psi (\mu )]=0  \label{3.14}
\end{equation}%
for all pairs $(\lambda ,\mu )\in \Sigma \times \Sigma _{k},$ $k=\overline{%
0,m}.$ This, in particular, means due to the Poincare lemma \cite{Go,Te}
that there exist differential $(k-1)$-forms $\Omega ^{(k-1)}[\varphi
(\lambda ),\psi (\mu )]\in \Lambda ^{k-1}(M;\mathbb{C}),$ $k=\overline{0,m},$
such that
\begin{equation}
Z^{(k)}[\varphi (\lambda ),\psi (\mu )]=d\Omega ^{(k-1)}[\varphi (\lambda
),\psi (\mu )]  \label{3.15}
\end{equation}%
for all pairs $(\varphi (\lambda ),\psi (\mu ))\in \mathcal{H}_{0}^{\ast
}\times \mathcal{H}_{\Lambda (\mathcal{L}),-}^{k}(M)$ parametrized by $\
(\lambda ,\mu )\in \Sigma \times \Sigma _{k},$ $k=\overline{0,m}.$ As a
result of passing on the right-hand side of (\ref{3.13}) to the homology
groups $H_{k}(M;\mathbb{C}),$ $k=\overline{0,m},$ one gets due to the
standard Stokes theorem \cite{Go} that the mappings
\begin{equation}
B_{\lambda }^{(k)}:\mathcal{H}_{\Lambda (\mathcal{L}),-}^{k}(M)%
\rightleftarrows H_{k}(M;\mathbb{C})  \label{3.16}
\end{equation}%
are isomorphisms for every $\lambda \in \Sigma .$ Making further use of the
Poincare duality \cite{Te} between the homology groups $H_{k}(M;\mathbb{C}),$
$k=\overline{0,m},$ and the cohomology groups $H^{k}(M;\mathbb{C}),$ $k=%
\overline{0,m},$ correspondingly, one obtains finally the statement claimed
in theorem 3.5, that is $\mathcal{H}_{\Lambda (\mathcal{L}),-}^{k}(M)\simeq
(H^{k}(M;\mathbb{C}))^{\Sigma }.\triangleright $
\end{proof}

\bigskip

Take now such a fixed pair $(\varphi (\lambda ),\psi (\mu ))\in \mathcal{H}%
_{0}^{\ast }\times \mathcal{H}_{\Lambda (\mathcal{L}),-}^{k}(M),$
parametrized by $(\lambda ,\mu )\in \Sigma \times \Sigma _{k},$ $k=\overline{%
0,m},$ for which due to both Theorem 3.3 and the \ Stokes theorem \cite%
{Go,Te} the equality
\begin{equation}
B_{\lambda }^{(k)}(\psi (\mu ))=S_{x}^{(k)}\int_{\partial S_{x}^{(k)}}\Omega
^{(k-1)}[\varphi (\lambda ),\psi (\mu )],  \label{3.17}
\end{equation}%
holds, where $S_{x}^{(k)}\in H_{k}(M;\mathbb{C}),$ $k=\overline{0,m},$ are
some arbitrary but fixed elements parametrized by an arbitrarily chosen
point $x\in M.$ Consider next the integral expressions
\begin{equation}
\Omega _{x}^{(k-1)}(\lambda ,\mu ):=\int_{\partial S_{x}^{(k)}}\Omega
^{(k-1)}[\varphi (\lambda ),\psi (\mu )],\text{ }\Omega
_{x_{0}}^{(k-1)}(\lambda ,\mu ):=\int_{\partial S_{x_{0}}^{(k)}}\Omega
^{(k-1)}[\varphi (\lambda ),\psi (\mu )]  \label{3.18}
\end{equation}%
and interpret them as the corresponding kernels \cite{Be} of the integral
invertible operators of Hilbert-Schmidt type $\ \Omega _{x}^{(k-1)},\Omega
_{x_{0}}^{(k-1)}:L_{2}^{(\rho )}(\Sigma ;\mathbb{C})\rightarrow L_{2}^{(\rho
_{k})}(\Sigma _{k};\mathbb{C}),$ $k=\overline{0,m},$ where $\rho $ and $\rho
_{k},$ $k=\overline{0,m},$ are some Lebesgue measures on the parameter sets $%
\Sigma $ and $\Sigma _{k},$ correspondingly. It assumes also above for
simplicity that boundaries $\partial S_{x}^{(k)}$ and $\partial
S_{x_{0}}^{(k)},$ $k=\overline{0,m},$ are taken homological to each other as
$x\rightarrow x_{0}\in M.$ Define now the expressions
\begin{equation}
\Omega ^{(k)}:\psi (\eta )\rightarrow \tilde{\psi}(\eta )  \label{3.19}
\end{equation}%
for $\psi (\eta )\in \mathcal{H}_{\Lambda (\mathcal{L}),-}^{k)}(M)$ and some
$\tilde{\psi}(\eta )\in \mathcal{H}_{\Lambda ,-}^{k)}(M),$ $k=\overline{0,m}%
, $ where, by definition
\begin{eqnarray}
\tilde{\psi}(\eta ) &:&=(\psi \Omega _{x}^{(k-1),-1}\Omega
_{x_{0}}^{(k-1)})(\eta )  \label{3.20} \\
&=&\int_{\Sigma _{k}}d\rho _{k}(\mu )\psi (\mu )\int_{\Sigma }d\rho (\xi
)\Omega _{x}^{(k-1),-1}(\mu ,\xi )\Omega _{x_{0}}^{(k-1)}(\xi ,\eta )  \notag
\end{eqnarray}%
for any $\eta \in \Sigma _{k},$ $k=\overline{0,m}.$

Suppose now that the elements (\ref{3.20}) are ones being related to some
another Delsarte transformed cohomology groups $\mathcal{H}_{\Lambda (%
\mathcal{\tilde{L}}),-}^{k}(M),$ $k=\overline{0,m},$ that is the following
condition
\begin{equation}
d_{\mathcal{\tilde{L}}}\tilde{\psi}(\eta )=0\text{ }  \label{3.21}
\end{equation}%
for $\tilde{\psi}(\eta )\in \mathcal{H}_{\Lambda (\mathcal{\tilde{L}}%
),-}^{k}(M),$ $\eta \in \Sigma _{k},$ $k=\overline{0,m},$ and some new
external anti-differentiation operation in $\mathcal{H}_{\Lambda ,-}(M)$%
\begin{equation}
d_{\mathcal{\tilde{L}}}:=\sum_{j=1}^{m}dx_{j}\wedge \tilde{L}_{j}(x;\partial
).  \label{3.22}
\end{equation}%
hold. Here, by definition, we will put
\begin{equation}
\tilde{L}_{j}:=\mathbf{\Omega }L_{j}\mathbf{\Omega }^{-1}  \label{3.23}
\end{equation}%
for each $\ j=\overline{1,m},$ where $\mathbf{\Omega }:\mathcal{H\rightarrow
}\mathcal{H}$ is the corresponding Delsarte transmutation operator. Since
all of operators $L_{j}:\mathcal{H\rightarrow }\mathcal{H},$ $j=\overline{1,m%
},$ were taken commuting, the same property also holds for the transformed
operators (\ref{3.23}), that is $[\tilde{L}_{j},\tilde{L}_{k}]=0,$ $k,j=%
\overline{0,m}.$ The latter is, evidently, equivalent due to (\ref{3.22}) to
the following general expression:%
\begin{equation}
d_{\mathcal{\tilde{L}}}=\mathbf{\Omega }d_{\mathcal{L}}\mathbf{\Omega }^{-1}.
\label{3.24}
\end{equation}%
For the condition (\ref{3.24}) and (\ref{3.21}) to be satisfied, let us
consider the corresponding to (\ref{3.17}) expressions
\begin{equation}
\tilde{B}_{\lambda }^{(k)}(\tilde{\psi}(\eta )=S_{x}^{(k)}\tilde{\Omega}%
_{x}^{(k-1)}(\lambda ,\eta )  \label{3.25}
\end{equation}%
related with the corresponding external differentiation (\ref{3.24}), where $%
S_{x}^{(k)}\in H_{k}(M;\mathbb{C})$ and $(\lambda ,\eta )\in \Sigma \times
\Sigma _{k},$ $k=\overline{0,m}.$ Assume further that there is also defined
a mapping
\begin{equation}
\mathbf{\Omega }^{\ast }\varphi (\lambda ):=\tilde{\varphi}(\lambda ),\text{
\ \ \ }\mathbf{\Omega }^{(k)}\psi (\eta ):=\tilde{\psi}(\eta ),  \label{3.26}
\end{equation}%
with $\mathbf{\Omega }^{\ast }:\mathcal{H}^{\ast }\mathcal{\rightarrow H}%
^{\ast }$ being an operator associated (but not necessary adjoint!) with the
basic Delsarte transmutation operator $\mathbf{\Omega }:\mathcal{%
H\rightarrow H}$ satisfying the standard relationships $\tilde{L}_{j}^{\ast
}:=\mathbf{\Omega }^{\ast }L_{j}^{\ast }\mathbf{\Omega }^{\ast ,-1},$ $j=%
\overline{1,m}.$ The corresponding Delsarte type operators $\mathbf{\Omega }%
^{(k)}:\mathcal{H}_{\Lambda (\mathcal{L}),-}^{k}(M)\rightarrow \mathcal{H}%
_{\Lambda (\mathcal{\tilde{L}}),-}^{k}(M),$\ $k=\overline{0,m},$ are related
with the action (\ref{3.20}) under the conditions \
\begin{equation}
d_{\mathcal{\tilde{L}}}\tilde{\psi}(\eta )=0,\text{ \ \ }d_{\mathcal{\tilde{L%
}}}^{\ast }\tilde{\varphi}(\lambda )=0,\text{ \ \ }  \label{3.27}
\end{equation}%
needed to be satisfied, meaning evidently that the elements $\tilde{\varphi}%
(\lambda )\in \mathcal{H}_{\Lambda (\mathcal{\tilde{L}}^{\ast }),-}^{0}(M),$
$\lambda \in \Sigma ,$ and elements $\tilde{\psi}(\eta )\in \mathcal{H}%
_{\Lambda (\mathcal{\tilde{L}}),-}^{k}(M),$ $\eta \in \Sigma _{k},$ $k=%
\overline{0,m}.$\ Now we need to formulate a lemma being important for the
conditions (\ref{3.27}) to hold.

\begin{lemma}
The following invariance property
\begin{equation}
\tilde{Z}^{(k)}=\Omega _{x_{0}}^{(k-1)}\Omega _{x}^{(k-1),-1}Z^{(k)}\Omega
_{x}^{(k-1),-1}\Omega _{x_{0}}^{(k-1)}  \label{3.28}
\end{equation}%
holds for any $k=\overline{0,m}.$
\end{lemma}

As a result of (\ref{3.28}) and the symmetry invariance between cohomology
spaces $\mathcal{H}_{\Lambda (\mathcal{L}),-}^{0}(M)$ and $\mathcal{H}%
_{\Lambda (\mathcal{\tilde{L}}),-}^{0}(M)$ one obtains the following pairs
of related mappings:
\begin{eqnarray}
\psi &=&\tilde{\psi}\tilde{\Omega}_{x}^{(k-1),-1}\tilde{\Omega}%
_{x_{0}}^{(k-1)},\text{ \ }\varphi =\tilde{\varphi}\tilde{\Omega}_{x}^{\ast
,-1}\tilde{\Omega}_{x_{0}},  \label{3.29} \\
\tilde{\psi} &=&\psi \Omega _{x}^{(k-1),-1}\Omega _{x_{0}}^{(k-1)},\text{ \
\ }\tilde{\varphi}=\varphi \Omega _{x}^{\ast ,-1}\Omega _{x_{0}}^{\ast },
\notag
\end{eqnarray}%
where the integral operator kernels defined as
\begin{eqnarray}
\Omega _{x}^{\ast }(\lambda ,\mu ) &:&=\int_{\partial S_{x}^{(m-1)}}\bar{%
\Omega}^{(m-2),\intercal }[\varphi (\lambda ),\psi (\mu )],\text{ }
\label{3.29a} \\
\tilde{\Omega}_{x}^{\ast }(\lambda ,\mu ) &:&=\int_{\partial S_{x}^{(m-1)}}%
\overset{\_}{\tilde{\Omega}}^{(m-2),\intercal }[\varphi (\lambda ),\psi (\mu
)]  \notag
\end{eqnarray}%
for all $(\lambda ,\eta )\in \Sigma \times \Sigma _{k},$ $k=\overline{0,m},$
giving rise to proper Delsarte transmutation operators ensuring the pure
differential nature of\ the transformed expressions (\ref{3.23}).

Note here also that due to (\ref{3.28}) and (\ref{3.29}) the following
operator property
\begin{equation}
\Omega _{x_{0}}^{(k-1)}\Omega _{x}^{(k-1),-1}\Omega _{x_{0}}^{(k-1)}+\tilde{%
\Omega}_{x_{0}}^{(k-1)}\Omega _{x}^{(k-1),-1}\Omega _{x_{0}}^{(k-1)}=0
\label{3.30}
\end{equation}%
holds for every $k=\overline{0,m},$ meaning that $\tilde{\Omega}%
_{x_{0}}^{(k-1)}=-\Omega _{x_{0}}^{(k-1)}.$

Take now $k=m-1;$ then one can define similar to (\ref{3.10}) the additional
closed and dense in $\mathcal{H}$ three subspaces%
\begin{equation}
\mathcal{H}_{0}:=\{\psi (\mu )\in \mathcal{H}_{\Lambda (\mathcal{L}%
),-}^{0}(M):d_{\mathcal{L}}\psi (\mu )=0,\text{ \ \ }\psi (\mu )|_{\Gamma
}=0,\text{ }\mu \in \Sigma \},  \notag
\end{equation}%
\begin{equation}
\mathcal{\tilde{H}}_{0}:=\{\tilde{\psi}(\mu )\in \mathcal{H}_{\Lambda (%
\widetilde{\mathcal{L}}),-}^{0}(M):d_{\widetilde{\mathcal{L}}}\tilde{\psi}%
(\mu )=0,\text{ \ \ }\tilde{\psi}(\mu )|_{\tilde{\Gamma}}=0,\text{ }\mu \in
\Sigma \},  \label{3.31}
\end{equation}%
\begin{equation*}
\mathcal{\tilde{H}}_{0}^{\ast }:=\{\tilde{\varphi}(\eta )\in \mathcal{H}%
_{\Lambda (\mathcal{\tilde{L}}^{\ast }),-}^{0}(M):d_{\widetilde{\mathcal{L}}%
}^{\ast }\tilde{\psi}(\eta )=0,\text{ \ \ }\tilde{\varphi}(\eta )|_{\tilde{%
\Gamma}}=0,\text{ }\eta \in \Sigma \},
\end{equation*}%
where $\Gamma $ and $\tilde{\Gamma}\subset M$ are some smooth $(m-2)$%
-dimensional hypersurfaces, and construct the actions
\begin{equation}
\mathbf{\Omega }:\psi \rightarrow \tilde{\psi}:=\psi \Omega _{x}^{-1}\Omega
_{x_{0}},\text{ \ \ \ }\mathbf{\Omega }^{\ast }:\varphi \rightarrow \tilde{%
\varphi}:=\varphi \Omega _{x}^{\ast ,-1}\Omega _{x_{0}}^{\ast }  \label{3.32}
\end{equation}%
on arbitrary but fixed pairs of elements $(\tilde{\varphi}(\lambda ),\tilde{%
\psi}(\mu ))\in \mathcal{H}_{0}^{\ast }\times \mathcal{H}_{0},$ parametrized
by the set $\Sigma ,$ where by definition, one needs that all obtained pairs
$(\tilde{\varphi}(\lambda ),\tilde{\psi}(\mu ))$ belong to $\mathcal{H}%
_{\Lambda (\mathcal{\tilde{L}}^{\ast }),-}^{0}(M)\times \mathcal{H}_{\Lambda
(\mathcal{\tilde{L}}),-}^{0}(M).$\bigskip\ Here for all $(\lambda ,\eta )\in
\Sigma \times \Sigma $ we defined, as usually,\ by expressions%
\begin{equation*}
\Omega _{x}(\lambda ,\mu ):=\int_{\partial S_{x}^{(m-1)}}\Omega
^{(m-2)}[\varphi (\lambda ),\psi (\mu )],\text{ \ }\Omega _{x}^{\ast
}(\lambda ,\mu ):=\int_{\partial S_{x}^{(m-1)}}\bar{\Omega}^{(m-2),\intercal
}[\varphi (\lambda ),\psi (\mu )],
\end{equation*}%
\begin{equation}
\Omega _{x_{0}}(\lambda ,\mu ):=\int_{\partial S_{x_{0}}^{(m-1)}}\Omega
^{(m-2)}[\varphi (\lambda ),\psi (\mu )],\ \Omega _{x_{0}}^{\ast }(\lambda
,\mu ):=\int_{\partial S_{x_{0}}^{(m-1)}}\bar{\Omega}^{(m-2),\intercal
}[\varphi (\lambda ),\psi (\mu )]  \label{3.33}
\end{equation}%
the corresponding kernels of integral operators acting in the Hilbert space $%
L_{2}^{(\rho )}(\Sigma ;\mathbb{C})$ of measurable functions with respect to
some Borel measure $\rho $ on Borel subsets of the set $\Sigma .$ The
related operator property (\ref{3.30}) can be compactly written down as
follows:%
\begin{equation}
\tilde{\Omega}_{x}=\tilde{\Omega}_{x_{0}}\Omega _{x}^{-1}\Omega
_{x_{0}}=-\Omega _{x_{0}}\Omega _{x}^{-1}\Omega _{x_{0}}.  \label{3.34}
\end{equation}

\bigskip Construct now from the expressions (\ref{3.33}) the following
operator quantities in the Hilbert space $L_{2}^{(\rho )}(\Sigma ;\mathbb{C}%
) $:%
\begin{equation}
\Omega _{x}-\Omega _{x_{0}}=\int_{\partial S_{x}^{(m-1)}}\Omega
^{(m-2)}[\varphi (\lambda ),\psi (\mu )]-\int_{\partial S_{x}^{(m-1)}}\Omega
^{(m-2)}[\varphi (\lambda ),\psi (\mu )]  \label{3.35}
\end{equation}%
\begin{equation*}
=\underset{S^{(m-1)}(\sigma _{x}^{(m-2)},\sigma _{x_{0}}^{(m-2)})}{\int }%
d\Omega ^{(m-2)}[\varphi (\lambda ),\psi (\mu )]=\underset{S^{(m-1)}(\sigma
_{x}^{(m-2)},\sigma _{x_{0}}^{(m-2)})}{\int }Z^{(m-1)}[\varphi (\lambda
),\psi (\mu )],
\end{equation*}%
\begin{equation*}
\Omega _{x}^{\ast }-\Omega _{x_{0}}^{\ast }=\int_{\partial S_{x}^{(m-1)}}%
\bar{\Omega}^{(m-2),\intercal }[\varphi (\lambda ),\psi (\mu
)]-\int_{\partial S_{x}^{(m-1)}}\bar{\Omega}^{(m-2),\intercal }[\varphi
(\lambda ),\psi (\mu )]
\end{equation*}%
\begin{equation*}
=\underset{S^{(m-1)}(\sigma _{x}^{(m-2)},\sigma _{x_{0}}^{(m-2)})}{\int }d%
\bar{\Omega}^{(m-2),\intercal }[\varphi (\lambda ),\psi (\mu )]=\underset{%
S^{(m-1)}(\sigma _{x}^{(m-2)},\sigma _{x_{0}}^{(m-2)})}{\int }%
Z^{(m-1),\intercal }[\varphi (\lambda ),\psi (\mu )],
\end{equation*}%
where, by definition, an $(m-1)$-dimensional surface $S^{(m-1)}(\sigma
_{x}^{(m-2)},\sigma _{x_{0}}^{(m-2)})\subset M$ is spanned smoothly between
two homological cycles $\sigma _{x}^{(m-2)}:=\partial S_{x}^{(m-1)}$ and $%
\sigma _{x_{0}}^{(m-2)}:=\partial S_{x_{0}}^{(m-1)}\in H_{m-1}(M;\mathbb{C}%
). $

Since the integral operator expressions $\Omega _{x_{0}},\Omega
_{x_{0}}^{\ast }:L_{2}^{(\rho )}(\Sigma ;\mathbb{C})\rightarrow L_{2}^{(\rho
)}(\Sigma ;\mathbb{C})$ are at a fixed point $x_{0}\in M$ evidently constant
and assumed to be invertible, for extending the actions given (\ref{3.32})
on the whole Hilbert space $\mathcal{H\times H}^{\ast }$ one can apply to
them the classical constants variation approach, making use of the
expressions (\ref{3.35}). As a result, we obtain easily the following
Delsarte transmutation integral operator expressions%
\begin{eqnarray}
\mathbf{\Omega } &=&\mathbf{1-}\int_{\Sigma \times \Sigma }d\rho (\xi )d\rho
(\eta )\tilde{\psi}(x;\xi )\Omega _{x_{0}}^{-1}[\tilde{\varphi}(\lambda ),%
\tilde{\psi}(\mu )](\xi |\eta )\times  \label{3.36} \\
&&\times \underset{S^{(m-1)}(\sigma _{x}^{(m-2)},\sigma _{x_{0}}^{(m-2)})}{%
\int }Z^{(m-1)}[\varphi (\eta ),\cdot ],  \notag
\end{eqnarray}%
\begin{eqnarray*}
\mathbf{\Omega }^{\ast } &=&\mathbf{1-}\int_{\Sigma \times \Sigma }d\rho
(\xi )d\rho (\eta )\tilde{\varphi}(x;\eta )\Omega _{x_{0}}^{\ast ,-1}[\tilde{%
\varphi}(\lambda ),\tilde{\psi}(\mu )](\eta |\xi )\times \\
&&\times \underset{S^{(m-1)}(\sigma _{x}^{(m-2)},\sigma _{x_{0}}^{(m-2)})}{%
\int }\text{ \ }\bar{Z}^{(m-1),\intercal }[\cdot ,\psi (\xi )]
\end{eqnarray*}%
for fixed pairs $(\tilde{\varphi}(\lambda ),\tilde{\psi}(\mu ))$ and $(%
\tilde{\varphi}(\lambda ),\tilde{\psi}(\mu ))\in \mathcal{\tilde{H}}%
_{0}^{\ast }\times \mathcal{\tilde{H}}_{0},$ being bounded invertible
integral operators of \ Volterra type on the whole space $\mathcal{H}\times
\mathcal{H}^{\ast }.$ Applying the same arguments as in Section 1, one can
show also that correspondingly transformed sets of operators $\tilde{L}_{j}:=%
\mathbf{\Omega }L_{j}\mathbf{\Omega }^{-1},$ $j=\overline{1,m},$ and \ $%
\tilde{L}_{k}^{\ast }:=\mathbf{\Omega }^{\ast }L_{k}^{\ast }\mathbf{\Omega }%
^{\ast ,-1},$ $k=\overline{1,m},$ appear to be purely differential too.
Thereby, one can formulate the following final theorem.

\begin{theorem}
The expressions (\ref{3.36}) are bounded invertible Delsarte transmutation
integral operators of Volterra type onto $\mathcal{H}\times \mathcal{H}%
^{\ast },$ transforming, correspondingly, given commuting sets of operators $%
L_{j},$ $j=\overline{1,m},$ and their formally adjoint ones $L_{k}^{\ast },$
$k$ $=\overline{1,m},$ into the pure differential sets of operators $\tilde{L%
}_{j}:=\mathbf{\Omega }L_{j}\mathbf{\Omega }^{-1},$ $j=\overline{1,m},$ and
\ $\tilde{L}_{k}^{\ast }:=\mathbf{\Omega }^{\ast }L_{k}^{\ast }\mathbf{%
\Omega }^{\ast -1},$ $k=\overline{1,m}.$ Moreover, the suitably constructed
closed subspaces $\mathcal{H}_{0}\subset \mathcal{H}$ \ and $\mathcal{\tilde{%
H}}_{0}\subset \mathcal{H},$ such that $\mathbf{\Omega }:\mathcal{H}%
_{0}\rightleftarrows \mathcal{\tilde{H}}_{0},$ depend strongly on the
topological structure of the generalized cohomology groups $\mathcal{H}%
_{\Lambda (\mathcal{L}),-}^{0}(M)$ and $\mathcal{H}_{\Lambda (\mathcal{%
\tilde{L}}),-}^{0}(M),$ parametrized by points $x,x_{0}\in M.$
\end{theorem}

\section{Discussion.}

Consider a differential operator $L:\mathcal{H\rightarrow H}$ in the form (%
\ref{1.1}) and assume that its spectrum $\sigma (L)$ consists of the
discrete $\sigma _{d}(L)$ and continuos $\sigma _{c}(L)$ parts. By means of
the general form of the Delsarte transmutation operators (\ref{3.36}) one
can construct a transformed more complicated differential operator $\tilde{L}%
:=\mathbf{\Omega }L\mathbf{\Omega }^{-1}$ in $\mathcal{H},$ such that its
continuous spectrum $\sigma _{c}(\tilde{L})=\sigma _{c}(L)$ but $\sigma
_{d}(L)\neq \sigma _{d}(\tilde{L}).$ Thereby these Delsarte transformed
operators can be effectively used for both studying spectral properties of
differential operators \cite{Be,Fa,Ma,LS} and constructing a wide class of
nontrivial differential operators with a prescribed spectrum as it was done
\-\cite{Ma,No} in one dimension.

As was shown before in \cite{Fa,Ni} for the two-dimensional Dirac and
three-dimensional perturbed Laplace operators, the kernels of the
corresponding Delsarte transmutation operator satisfy some special of
Fredholm type linear integral equations called the Gelfand-Levitan-Marchenko
ones, which are of very importance for solving the corresponding inverse
spectral problem and having many applications in modern mathematical
physics. Such equations can be naturally constructed for our
multidimensional case too, thereby making it possible to pose the
corresponding inverse spectral problem for describing a wide class of
multidimensional operators with a priori given spectral characteristics. The
mentioned problem appears (see \cite{Be}) to be strongly related with that
of spectral representation of kernels commuting in some sense with a given
pair of differential operators. Also, similar to \cite{Ni,PM}, one can use
such results for studying so called completely integrable nonlinear
evolution equations, especially for constructing by means of special Darboux
transformations \cite{MS,SPS} their exact solutions like solitons and many
others. Such an activity is now in progress and the corresponding results
will be published later.

\section{Acknowledgements.}

One of authors (A.P.) cordially thanks prof. L.P. Nizhnik (IM of NAS, Kyiv,
Ukraina), prof. T. Winiarska (IM, Politechnical University of Krakow,
Poland), profs. A. Pelczar and J. Ombach (UJ, Krakow), prof. J. Janas (IM of
PAN, Krakow, Poland), prof. Z. Peradzynski (UW, Warszaw) and prof. D.L.
Blackmore (NJIT, Newark, NJ, USA) for valuable discussions of the problem
studied in the article.

\bigskip

\end{document}